# JGR Space Physics

AGU ADVANCING EARTH AND SPACE SCIENCES

# Whistler Waves in the Quasi-Parallel and Quasi-Perpendicular Magnetosheath

I. Svenningsson[1,2] , E. Yordanova[1] , Yu. V. Khotyaintsev[1,2] , M. André[1] , G. Cozzani[3] , and K. Steinvall[4]

[1]Swedish Institute of Space Physics, Uppsala, Sweden, [2]Department of Physics and Astronomy, Uppsala University, Uppsala, Sweden, [3]Department of Physics, University of Helsinki, Helsinki, Finland, [4]School of Physics and Astronomy, University of Southampton, Southampton, UK

**Abstract** In the Earth's magnetosheath (MSH), several processes contribute to energy dissipation and plasma heating, one of which is wave-particle interactions between whistler waves and electrons. However, the overall impact of whistlers on electron dynamics in the MSH remains to be quantified. We analyze 18 hr of burst-mode measurements from the Magnetospheric Multiscale (MMS) mission, including data from the unbiased magnetosheath campaign during February–March 2023. We present a statistical study of 34,409 whistler waves found using automatic detection. We compare wave occurrence in the different MSH geometries and find three times higher occurrence in the quasi-perpendicular MSH compared to the quasi-parallel case. We also study the wave properties and find that the waves propagate quasi-parallel to the background magnetic field, have a median frequency of 0.2 times the electron cyclotron frequency, median amplitude of 0.03–0.06 nT (30–60 pT), and median duration of a few tens of wave periods. The whistler waves are preferentially observed in local magnetic dips and density peaks and are not associated with an increased temperature anisotropy. Also, almost no whistlers are observed in regions with parallel electron plasma beta lower than 0.1. Importantly, when estimating pitch-angle diffusion times we find that the whistler waves cause significant pitch-angle scattering of electrons in the MSH.

## 1. Introduction

The Earth's magnetosheath (MSH) is a dynamic and variable region with properties that depend strongly on the upstream conditions, in particular the shock normal angle (the angle between the interplanetary magnetic field and the bow shock normal), $\theta_{Bn}$. The quasi-parallel ($Q_\parallel$; $\theta_{Bn} < 45°$) MSH is generally more variable than the quasi-perpendicular ($Q_\perp$; $\theta_{Bn} > 45°$) MSH. The $Q_\parallel$ MSH is influenced by the ion foreshock which results in stronger magnetic fluctuations (Luhmann et al., 1986; Turc et al., 2023). The $Q_\parallel$ MSH also contains high-energy ions accelerated by the $Q_\parallel$ bow shock (Fuselier et al., 1991; Johlander et al., 2021). In the $Q_\perp$ MSH, on the other hand, there is a high ion temperature anisotropy which can excite mirror mode waves (Dimmock et al., 2015).

In the magnetosheath, several processes such as turbulence and wave-particle interactions contribute to collisionless energy transfer and dissipation. One wave known to interact with electrons is the whistler wave, or whistler, a right-hand polarized wave with frequency below the electron cyclotron frequency $f_{ce} = |q_e|B/(2\pi m_e)$ (where $q_e$ is the electron charge, $B$ the magnetic field strength, and $m_e$ the electron mass). Whistlers are frequently observed in the Earth's MSH (see examples below) as well as in the solar wind (Cattell et al., 2021; Lin et al., 1995; Stone et al., 1995), the radiation belts (Li et al., 2017), the magnetotail (Huang et al., 2017; Khotyaintsev et al., 2011), and the dayside magnetopause (Vaivads et al., 2007). In the MSH, the strongest whistler emissions have historically been termed *lion roars* (Smith & Tsurutani, 1976).

Whistler waves can be generated by various instabilities. The most commonly used example is the electron temperature anisotropy instability in a bi-Maxwellian plasma. It is triggered when the perpendicular electron temperature ($T_{e\perp}$) sufficiently exceeds the parallel temperature $T_{e\parallel}$ ($\perp$ and $\parallel$ defined with respect to the background magnetic field) (Kennel & Petschek, 1966). This instability has been observed in the MSH (Huang et al., 2018). However, since electron distribution functions in the MSH are often non-Maxwellian, other generation mechanisms can generate whistler waves without requiring a high temperature anisotropy. One such unstable non-Maxwellian electron distribution is the so-called butterfly distribution, which is characterized by phase-space density minima at pitch angles 0, 90, and 180°. Such distributions have recently been shown to generate whistler waves in mirror modes (Kitamura et al., 2020), magnetic holes (H. Zhang et al., 2021), and in local magnetic fluctuations downstream of a $Q_\parallel$ bow shock (Svenningsson et al., 2022). Other examples are distributions with a high temperature anisotropy in a particular energy range (Breuillard, Le Contel et al., 2018) and bi-directional, anisotropic beams (Huang et al., 2020). Several of the above-cited studies solved the dispersion relation, confirming that the local electron velocity distributions were unstable to the observed waves (Huang et al., 2018; Huang et al., 2020; H. Zhang et al., 2021; Svenningsson et al., 2022). Common observed wave properties for these case studies are parallel propagation to the background magnetic field and frequency in the range 0.05–0.25 $f_{ce}$.

The above-mentioned whistler observations occurred in local minima of $B$, henceforth called *magnetic dips*. In the case of mirror modes, this was accompanied by local maxima in the density $n$, or *density peaks*. The correlation between magnetic dips and whistler waves has been known for a long time (e.g., Smith & Tsurutani, 1976), but waves are also frequently observed outside magnetic dips (Y. Zhang et al., 1998). There are physical arguments as to why magnetic dips and density peaks can constitute source regions for whistler waves. A local decrease in $B$ and an increase in $n$ lead to a decrease in resonant energy (Kennel & Petschek, 1966), which increases the number of electrons that can interact with the wave. In addition, a local $B$ minimum traps particles above a critical pitch angle $\theta_{tr}$ defined through $\sin\theta_{tr} = \sqrt{B/B_{max}}$ (Yao et al., 2018). This can create electron deficits in the parallel/anti-parallel directions which are unstable to whistler wave generation (Kitamura et al., 2020; Svenningsson et al., 2022).

Besides the case studies mentioned above, there have been several statistical studies of whistler waves in the MSH (Smith & Tsurutani, 1976; Rodriguez, 1985; Y. Zhang et al., 1998; Baumjohann et al., 1999; Yearby et al., 2005; Giagkiozis et al., 2018). These studies found average frequencies of 0.1–0.3$f_{ce}$ and quasi-parallel propagation with respect to the background magnetic field (wave normal angle $\theta_k < 30°$). All of them considered the MSH region as a whole without distinguishing between $Q_\parallel$ and $Q_\perp$ MSH. The exception is Rodriguez (1985), which focused specifically on intense, long-duration (>5 s) lion roars (LDLRs) and found that they were almost exclusively found downstream of $Q_\perp$ shocks. However, LDLRs are a minority since the observed average whistler duration ranges between 0.07 s (Yearby et al., 2005) and 1.6 s (Smith & Tsurutani, 1976). Since mirror modes (whistler source regions as described above) are typical of the $Q_\perp$ MSH, whistler wave activity is expected in this geometry. However, apart from a case study of one MSH interval (Svenningsson et al., 2022), it has not been investigated to what extent whistler waves can form in the more turbulent $Q_\parallel$ MSH geometry.

Electrons in resonance with whistler waves change their energy and pitch angle, resulting in a reshaping of the electron velocity distribution function. Shi et al. (2020) showed that whistlers at the bow shock effectively scatter electrons and modify the electron distribution. In a recent numerical study, Jiang et al. (2023) modeled a MSH electron distribution during the excitation of whistler waves. They observed that energy transfer mainly occurred through the first-order cyclotron resonance and that the interaction caused a reduction of the electron temperature anisotropy. However, the effect on the electron distribution from whistler waves in the MSH has not been investigated statistically, and the question of whether the whistlers have a major impact on the MSH electron dynamics remains open.

In this study, we analyze 18 hr of high-resolution measurements from the MSH measured by the Magnetospheric Multiscale (MMS) mission (Burch et al., 2016). Compared to previous statistical studies, we separate the waves into the $Q_\perp$ and $Q_\parallel$ MSH. In addition, we study and compare the wave properties in the two geometries. We also estimate pitch-angle diffusion coefficients to assess whether whistler waves are significant to the electron dynamics in the MSH, which has not been done before.

In Section 2, we describe how we classify the MSH intervals and identify whistler wave packets. In Section 3, we compare whistler occurrence and wave properties in the two MSH geometries. We also investigate which conditions in the MSH favor the occurrence of whistler waves. In Section 4 we discuss the statistical results and estimate the effect of pitch-angle scattering from the waves. We also discuss a reduced whistler occurrence during periods of low plasma beta by investigating the dependence of the resonant energy on plasma beta and normalized wave frequency. Finally, we present our conclusions in Section 5.

## 2. Methods

In this study, we use burst-mode measurements from MMS. The Fast Plasma Investigation (FPI) (Pollock et al., 2016) instruments provide 3D velocity distributions and the derived moments with a time resolution of 150 and 30 ms for ions and electrons, respectively. The background magnetic field vector $\boldsymbol{B}$ is measured with 128 samples/s by the flux-gate magnetometer (FGM) (Russell et al., 2016). Wave measurements with 8,192 samples/s are made by the search coil magnetometer (SCM) (Le Contel et al., 2016) and the electric field double probe (EDP) (Ergun et al., 2016; Lindqvist et al., 2016). We also use the lower-resolution fast-mode data for overview plots in Section 2.1 and background magnetic field and ion density for the statistics in Section 3.3.

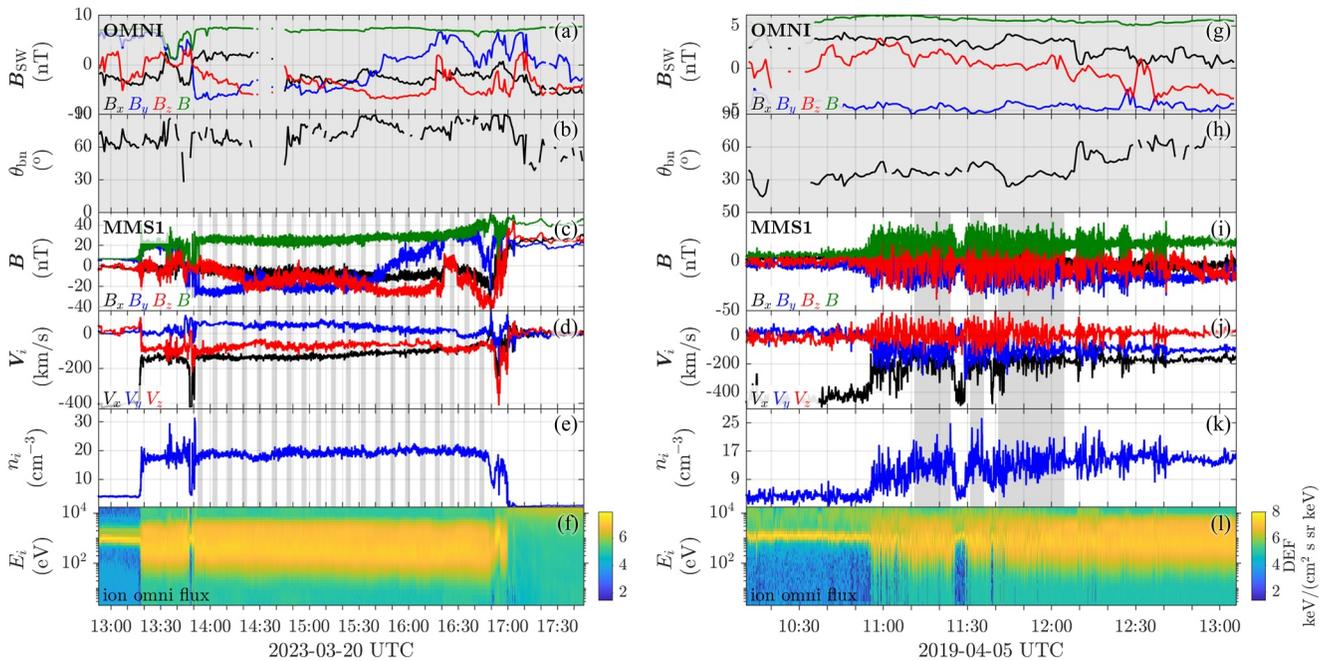

**Figure 1.** Examples of MSH intervals used in this study. Panels (a)–(f) show a $Q_\perp$ interval with $\langle\theta_{Bn}\rangle = 85°$. (a) Solar wind magnetic field $\boldsymbol{B}_{SW}$ propagated to the bow shock. (b) Shock normal angle $\theta_{Bn}$. (c) Magnetic field $\boldsymbol{B}$. (d) Ion velocity $\boldsymbol{V}_i$. (e) Ion density $n_i$. (f) Ion omnidirectional differential energy flux. Panels (a)–(b) show OMNI data and (c)–(f) show MMS data. (g)–(l): Same as (a)–(f) but for a $Q_\parallel$ interval with $\langle\theta_{Bn}\rangle = 35°$. The shaded regions in panels (c)–(e) and (i)–(k) show burst data intervals in which we searched for whistler waves. Here and throughout the paper, times are given in UTC.

In the next section, we describe the identification and classification of the MSH intervals and the detection of whistler waves in these intervals using an automated search routine.

## 2.1. Magnetosheath Classification

To define our $Q_\perp$ and $Q_\parallel$ MSH data sets, we use the shock normal angle $\theta_{Bn}$ calculated from a model described below, which is then verified by visual inspection.

MMS burst data intervals are selected manually by a *scientist in the loop* (SITL) (Fuselier et al., 2016). The burst selection naturally targets intervals of particular interest, such as sharp boundaries, jets, and magnetic reconnection sites, rather than undisturbed, typical MSH conditions. When performing statistical studies on MMS data, we need to be aware of the possible bias introduced by this manual selection.

For the $Q_\perp$ MSH, we used data from the recent *unbiased magnetosheath campaign*: To avoid the burst selection bias, three-minute intervals of burst data were collected every 9 min for 14 inbound MSH crossings during February–March 2023. One such interval is shown in Figures 1a–1f, with the 3-min burst intervals shaded gray. Panel a shows the upstream magnetic field $\boldsymbol{B}_{SW}$ from the OMNI database (Papitashvili & King, 2020), measured by spacecraft positioned at the first Lagrangian point and propagated to the bow shock nose. We calculated the shock normal angle $\theta_{Bn}$ (panel b) using a parabolic bow shock model (Merka et al., 2003):

$$X_{GSE} = a_s - b_s (Y_{GSE}^2 + Z_{GSE}^2),\tag{1}$$

where $(X_{GSE}, Y_{GSE}, Z_{GSE})$ are the GSE (Geocentric Solar Ecliptic) coordinates. The bow shock standoff distance $a_s$ (Farris & Russell, 1994) and the flaring parameter $b_s$ (Fairfield et al., 2001) depend on upstream solar wind parameters taken from the OMNI database (Papitashvili & King, 2020). We traced the spacecraft location along $X_{GSE}$ to the surface defined by Equation 1 and calculated the model shock normal direction $\hat{n}$ at that point. We then calculated $\theta_{Bn}$ as the angle between $\hat{n}$ and $B_{SW}$. In panel b, we see that $\theta_{Bn} > 45°$, indicating a $Q_\perp$ MSH. We also verified visually that the local MSH properties corresponded to typical $Q_\perp$ MSH characteristics: few rotations in $B$ (panel c), and relatively small fluctuations in $B$ (Karlsson et al., 2021) and plasma parameters such as the ion velocity $V_i$ (panel d) and density $n_i$ (panel e). The omnidirectional ion energy flux (panel f) is used to differentiate the MSH plasma (warm, broad energy distribution) from the solar wind (monoenergetic beam seen before 13:20) and magnetosphere (high-energy ions visible after 17:00).

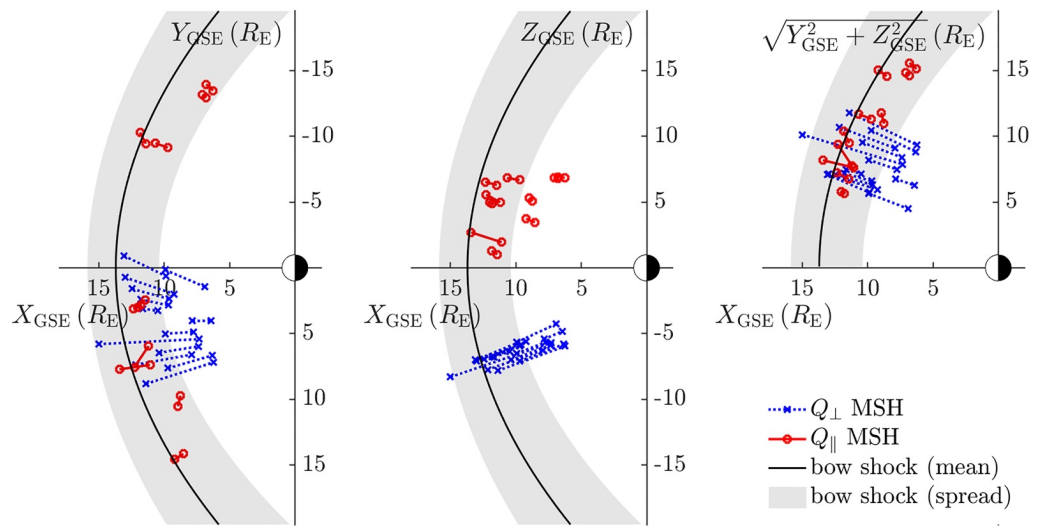

**Figure 2.** MMS locations during the MSH intervals included in our data set. Blue symbols connected with dotted lines mark the start and end of the $Q_\perp$ intervals MSH and the red ones show the same for the $Q_\parallel$ MSH. The black curve shows the average bow shock location (Equation 1) and the gray region shows the bow shock location spread among the intervals. Here, $1\,R_E = 6{,}371$ km is the Earth's radius.

Unfortunately, there was minimal coverage of the $Q_\parallel$ MSH during the unbiased campaign. Therefore, we complemented the data set of the 2023 campaign with intervals from 2018 to 2020 selected specifically to study magnetosheath turbulence, without aiming for particular structures or phenomena. Figures 1g–1l shows an example (same panels as a–f). Here, $\theta_{Bn} < 45°$ (panel h) for the duration of the burst intervals (shaded gray intervals in panels i–k), indicating a $Q_\parallel$ MSH. We also observe strong fluctuations and rotations in $B$ (panel i) and fluctuating $V_i$ and $n_i$ (panels j–k), contrasting the conditions in the $Q_\perp$ MSH.

Following the above classification procedure, we created $Q_\parallel$ and $Q_\perp$ data sets using the criterion of mean $\theta_{Bn} < 45°$ and $\theta_{Bn} > 45°$, respectively. The locations of the intervals are shown in Figure 2; blue showing $Q_\perp$ MSH and red showing $Q_\parallel$ MSH. The black curve shows the average bow shock location (Equation 1) and the gray region shows the bow shock location spread among the intervals. A table summarizing our MSH data set is given in Table S1 in Supporting Information S1.

### 2.2. Whistler Wave Detection

We divide each MSH interval from our data sets into sub-intervals of 10 s and search for whistler waves by calculating wavelet spectra of the fluctuating fields. We define the background $P(f)$ as the magnetic power spectral density $P(t, f)$ averaged over the 10-s sub-interval. Polarization parameters ($\theta_k$, ellipticity, degree of polarization, and planarity) are computed by the singular value decomposition method (Santolík et al., 2003). To detect whistler waves, we form $P^w(t, f)$, that is, $P(t, f)$ including only points satisfying the *whistler criteria*: high planarity and degree of polarization (>0.7), and high ellipticity (>0.7, corresponding to right-hand polarization). We also restrict the frequency range to $(0.05 f_{ce}, f_{ce})$. We then identify whistler wave packets using an approach similar to Bortnik et al. (2007): at every point in time (corresponding to the wavelet time resolution of 5 ms), spectral peaks are identified with power two times higher than the background ($P^w(t, f) \geq 2P(f)$). For each spectral peak, we identify the frequency with the highest power ($f_{peak}(t)$), and extract the maximum power $P^w_{max}(t)$ and the wave normal angle $\theta_k(t)$ at $f_{peak}(t)$. We then go through the spectral peaks chronologically and group them into wave packets; in other words, for each spectral peak, we determine if it belongs to the same wave packet as previous peaks, or if it is the first peak in the next wave packet. The association is based on the wave period $\tau^*_{wave} = 1/f^*_{peak}$, where $f^*_{peak}$ is the frequency calculated from the previous spectral peaks in the wave packet. We apply the following criteria:

1. If the time between two spectral peaks is less than $2\tau^*_{wave}$, they are grouped into the same packet.
2. Wave packets with durations less than $8\tau^*_{wave}$ are discarded.

Once the wave packets have been defined, we calculate the following wave properties: the duration $\Delta t = t_{end} - t_{start}$, where $t_{start}$ and $t_{end}$ are the start and end of the wave packet; the frequency $f = \langle f_{peak}(t) \rangle$ and the wave normal angle $\theta_k = \langle \theta_k(t) \rangle_P$, where $\langle \ldots \rangle_P$ denotes the time average weighted by $P^w_{max}(t)$. We estimate the amplitude $\delta B = \sqrt{2P^w_{max} \Delta f}$ where $P^w_{max} = \max(P^w_{max}(t))$ and $\Delta f$ the frequency bin width at $f$. We investigate these wave properties statistically in Section 3.1.

Figure 3 shows an example of an interval analyzed with this routine. Panel a shows the background magnetic field $\boldsymbol{B}$ and panel b shows the fluctuating field $\delta B$ in field-aligned coordinates, filtered above 25 Hz (the minimum value of $0.05f_{ce}$ in the interval). The $\delta B$ fluctuations correspond to the $P(t, f)$ enhancements in panel c. These are right-hand polarized waves (ellipticity = +1; panel d) that propagate quasi-parallel to the background magnetic field ($\theta_k < 30°$; panel e). In panel c, the black rectangles show the identified whistler wave packets, spanning from $t_{start}$ to $t_{end}$ in time. The frequency $f$ of the wave packet is indicated with a thinner black line and the top and bottom sides of the rectangle mark the highest and lowest frequencies containing whistler wave power, respectively. In Figures 3g and 3h, we show the magnetic waveforms in field-aligned coordinates for two of the wave packets. The right-handed field-aligned coordinate system $(\hat{e}_\parallel, \hat{e}_{\perp 1}, \hat{e}_{\perp 2})$ is aligned with $\boldsymbol{B}$ (panel a) measured by the FGM, with $\hat{e}_\parallel \| \boldsymbol{B}$ and $\hat{e}_{\perp 1} \| \boldsymbol{B} \times (1, 0, 0)_{GSE}$. We observe fluctuations in all components (panel g–h), but $\delta B_{\perp 1,2}$ are significantly larger than $\delta B_\parallel$ consistent with quasi-parallel propagation. Due to the variable conditions in the MSH, the waveforms can have different characteristics; for example, the wave packet in panel g has a high and variable amplitude ($\sim 0.4$ nT) similar to the lion roars studied by Giagkiozis et al. (2018). Panel f shows the parallel electron beta $\beta_{e\parallel} = 2\mu_0 n_e T_{e\parallel}/B^2$ (blue) and electron temperature anisotropy $T_{e\perp}/T_{e\parallel}$ (black, solid). Here $\mu_0$ is vacuum permeability and $n_e$ is the electron density. The black dashed curve shows the whistler instability threshold $T_{e\perp}/T_{e\parallel} = 1 + 0.36\beta_{e\parallel}^{-0.55}$ (assuming a bi-Maxwellian electron distribution), corresponding to the growth rate $\gamma = 0.01|\Omega_{ce}|$ (Gary & Wang, 1996), where $\Omega_{ce} = q_e B/m_e$ is the angular electron cyclotron frequency. We note that $T_{e\perp}/T_{e\parallel} \approx 1$ and is well below the threshold (predicting the plasma to be stable) during the whistler observations.

## 3. Results

### 3.1. Whistler Occurrence in the Quasi-Parallel and Quasi-Perpendicular Magnetosheath

We identified 34,409 whistler wave packets using the automatic search routine described above. We now compare the whistler occurrence rate in the $Q_\parallel$ MSH with those in the $Q_\perp$ MSH. In the $Q_\perp$ MSH, we detected more waves (41 per min) than in the $Q_\parallel$ MSH (14 per min) (see Table 1). We also calculate the *filling factor*, defined as

$$\text{filling factor} = \frac{\sum \Delta t(\text{wave packets})}{\sum \Delta t(\text{MSH intervals})}, \tag{2}$$

which quantifies the fraction of a MSH interval occupied by whistler waves. On average, we find that this fraction is higher in the $Q_\perp$ than in the $Q_\parallel$ MSH (17% compared to 6%). In Table S1 in Supporting Information S1, we give these numbers for each MSH interval. Based on the number of waves per minute and the filling factor, we conclude that the whistler waves are three times more common in the $Q_\perp$ MSH.

We now compare the wave properties in the different regions. In Figure 4, we show properties for the whistler waves observed in the $Q_\perp$ (blue) and $Q_\parallel$ (red) MSH. Figure 4a shows the wave normal angle $\theta_k$. We see that the whistlers are predominantly parallel propagating with median $\theta_k = 12°$ ($Q_\perp$ MSH) and $\theta_k = 17°$ ($Q_\parallel$ MSH) (medians provided in the legend), in agreement with previous studies (Giagkiozis et al., 2018). We observe a slightly larger median $\theta_k$, and a more pronounced tail of oblique whistlers, in the $Q_\parallel$ MSH. In Figure 4b, we show $f/f_{ce}$. The median is similar in both geometries ($f/f_{ce} = 0.23$ and $0.21$ in the $Q_\perp$ and $Q_\parallel$ MSH, respectively), but there is a larger spread of frequencies in the $Q_\parallel$ MSH. Panel c shows the frequency $f$ in Hz. The median $f$ is higher (170 Hz) in the $Q_\perp$ than in the $Q_\parallel$ MSH (109 Hz), due to an on-average stronger magnetic field in this data set (see Table S1 in Supporting Information S1). The median whistler amplitude $\delta B$ is 0.03 nT in the $Q_\perp$ MSH and 0.06 nT in the $Q_\parallel$ MSH (see Figure 4d). The median wave duration $\Delta t$ is a few tens of wave periods; $21\tau_{\text{wave}}$ (0.13 s) and $17\tau_{\text{wave}}$ (0.17 s) in the $Q_\perp$ and $Q_\parallel$ MSH, respectively (see Figures 4e and 4f).

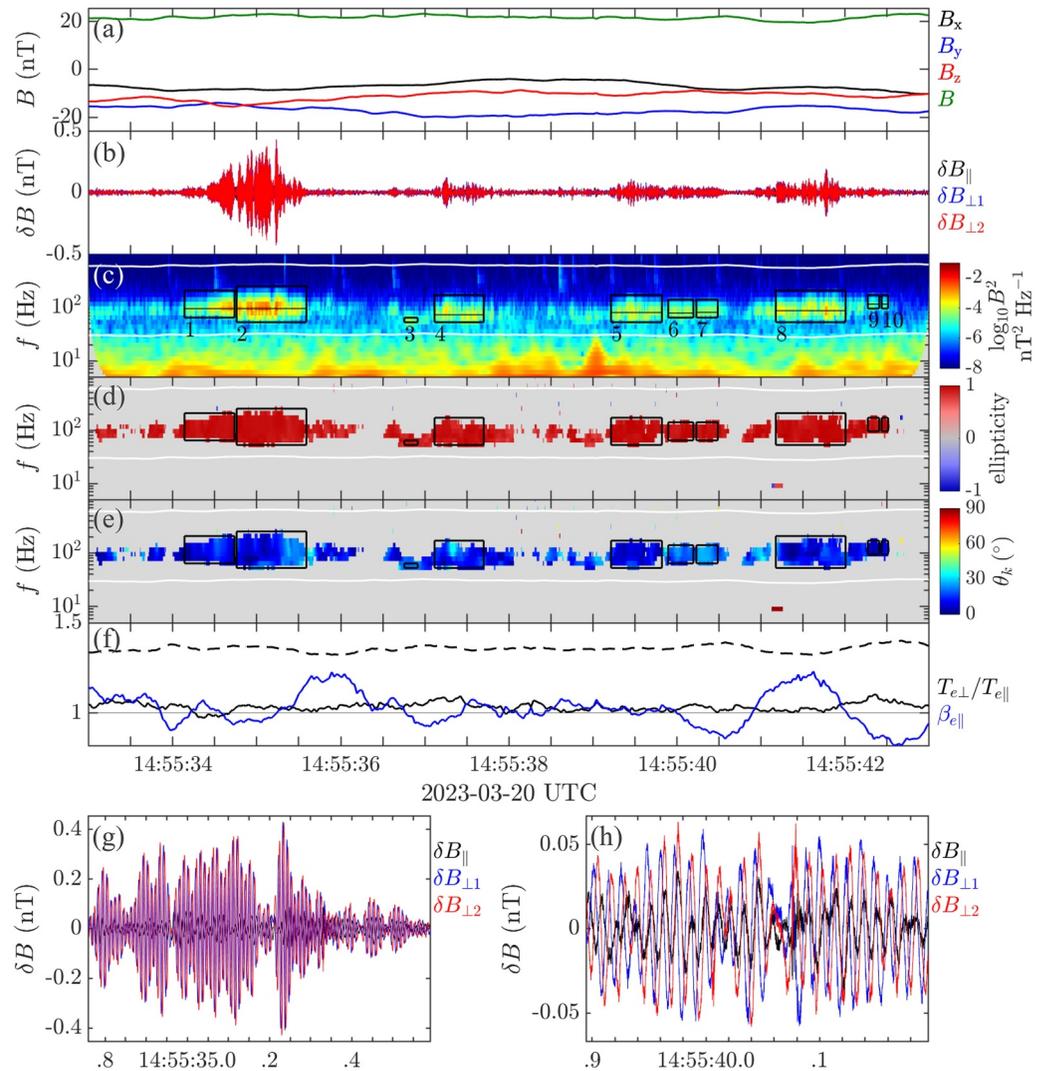

**Figure 3.** Examples of wave packets found with the wave detection routine. (a) Background magnetic field $\boldsymbol{B}$ in GSE coordinates. (b) Magnetic fluctuations $\delta\boldsymbol{B}$ in field-aligned coordinates, filtered above 25 Hz. (c) Magnetic field power spectrum $P(t, f)$. (d) Ellipticity. (e) Wave normal angle $\theta_k$. In (d)–(e), we only include points satisfying the whistler criteria (planarity, degree of polarization, and ellipticity >0.7). The white lines correspond to the frequencies $0.05f_{ce}$ and $f_{ce}$. (f) $\beta_{e\parallel}$ (blue), $T_{e\perp}/T_{e\parallel}$ (black, solid), whistler instability threshold for $\gamma = 0.01|\Omega_{ce}|$ (black, dashed). (g),(h) Magnetic field components in field-aligned coordinates (packets 2 and 6 in panel c).

In Figures 5a and 5b, we show 2D histograms of some properties of our wave packets. The solid curves show the median and the dashed curves the 25th–75th percentiles. Panel a shows a weak trend of increasing $\theta_k$ with higher $f/f_{ce}$; however, the difference is negligible for $f/f_{ce} < 0.5$. We observe that $\delta B$ drops with increasing $f/f_{ce}$ (panel b). When investigating the time average of $T_{e\perp}/T_{e\parallel}$ and $\beta_{e\parallel}$ during each wave packet, we find that higher-frequency waves ($f/f_{ce} \gtrsim 0.3$) are found in regions with slightly higher $T_{e\perp}/T_{e\parallel}$ (panel c) and lower $\beta_{e\parallel}$ (panel d) than the lower-frequency waves. We averaged $T_{e\perp}/T_{e\parallel}$ and $\beta_{e\parallel}$ without any weighting on for example, wave power and in the few cases where the wave packet was shorter than the electron data resolution of 30 ms, we used the value closest in time to the wave packet. We analyze the $\beta_{e\parallel}$-$T_{e\perp}/T_{e\parallel}$ parameter space further in Section 3.2. In Figure 5, we combine wave packets from the $Q_\perp$ and $Q_\parallel$ geometries; however, we have verified that the same qualitative trends appear in both data sets.

**Table 1**
*Whistler Occurrence and Wave Properties (Median Values; Histograms Presented in Figure 4) in the $Q_\perp$ and $Q_\parallel$ MSH, Respectively*

| MSH geometry | $Q_\perp$ | $Q_\parallel$ |
|---|---|---|
| Total burst time | 12 hr | 5.5 hr |
| Number of wave packets | 29,849 | 4,560 |
| Wave packets per minute | 41 | 14 |
| Filling factor | 17% | 6% |
| Wave normal angle $\theta_k$ | 12° | 17° |
| Frequency $f$ | 170 Hz or $0.23 f_{ce}$ | 109 Hz or $0.21 f_{ce}$ |
| Amplitude $\delta B$ | 0.03 nT or 0.001B | 0.06 nT or 0.004B |
| Duration $\Delta t$ | 0.13 s or $21\tau_{wave}$ | 0.17 s or $17\tau_{wave}$ |

In summary, we find that whistler waves are approximately three times more common in the $Q_\perp$ MSH than in the $Q_\parallel$. The wave properties are roughly similar in the two geometries: the waves are quasi-parallel to the background magnetic field (median $\theta_k = 12°$ and $17°$ in the $Q_\perp$ and $Q_\parallel$ MSH, respectively), the frequency is around $0.2 f_{ce}$, and the median duration a few to tens of wave periods. The frequency and quasi-parallel propagation confirm previous statistical studies (Baumjohann et al., 1999; Giagkiozis et al., 2018; Rodriguez, 1985; Smith & Tsurutani, 1976; Yearby et al., 2005; Y. Zhang et al., 1998). The whistlers in the $Q_\parallel$ MSH have a slightly higher median amplitude (0.06 nT) than in the $Q_\perp$ MSH (0.03 nT). The median $\theta_k$, $f$, $\delta B$, and $\Delta t$ values are included in Table 1.

### 3.2. Distribution of Observations in the $\beta_{e\parallel}$-$T_{e\perp}/T_{e\parallel}$ Parameter Space

Now we investigate the distribution of wave packets in the $\beta_{e\parallel}$-$T_{e\perp}/T_{e\parallel}$ parameter space, see Figure 6. This allows for comparison with relevant instability thresholds such as the whistler temperature anisotropy instability, assuming bi-Maxwellian electron distributions (Gary & Wang, 1996). Panels b and c show the histograms of $\beta_{e\parallel}$ and $T_{e\perp}/T_{e\parallel}$ for the $Q_\perp$ MSH data set; the filled histogram corresponds to the whistler wave packets and the histogram in solid black line to the whole $Q_\perp$ data set. First, we note that in the $Q_\perp$ MSH, the median $\beta_{e\parallel}$ (panel b) was higher during whistler wave observations (0.9) than it was for the whole data set (0.6). Interestingly, our data set includes periods of low $\beta_{e\parallel}$ which contain very few whistler waves. This low-$\beta_{e\parallel}$ tail comes from intervals with strong magnetic field and parallel electron temperature anisotropy during which the MSH was impacted by disturbed solar wind structures such as coronal mass ejections and stream interaction regions (see notes in Table S1 in Supporting Information S1). From Figure 6c, we see that the median $T_{e\perp}/T_{e\parallel} \approx 1$ during whistler wave observations is almost the same as for the whole $Q_\perp$ data set, that is, that the presence of whistler waves is not associated with an increased $T_{e\perp}/T_{e\parallel}$. However, for $T_{e\perp}/T_{e\parallel} \lesssim 0.8$ the occurrence of whistlers decreases significantly.

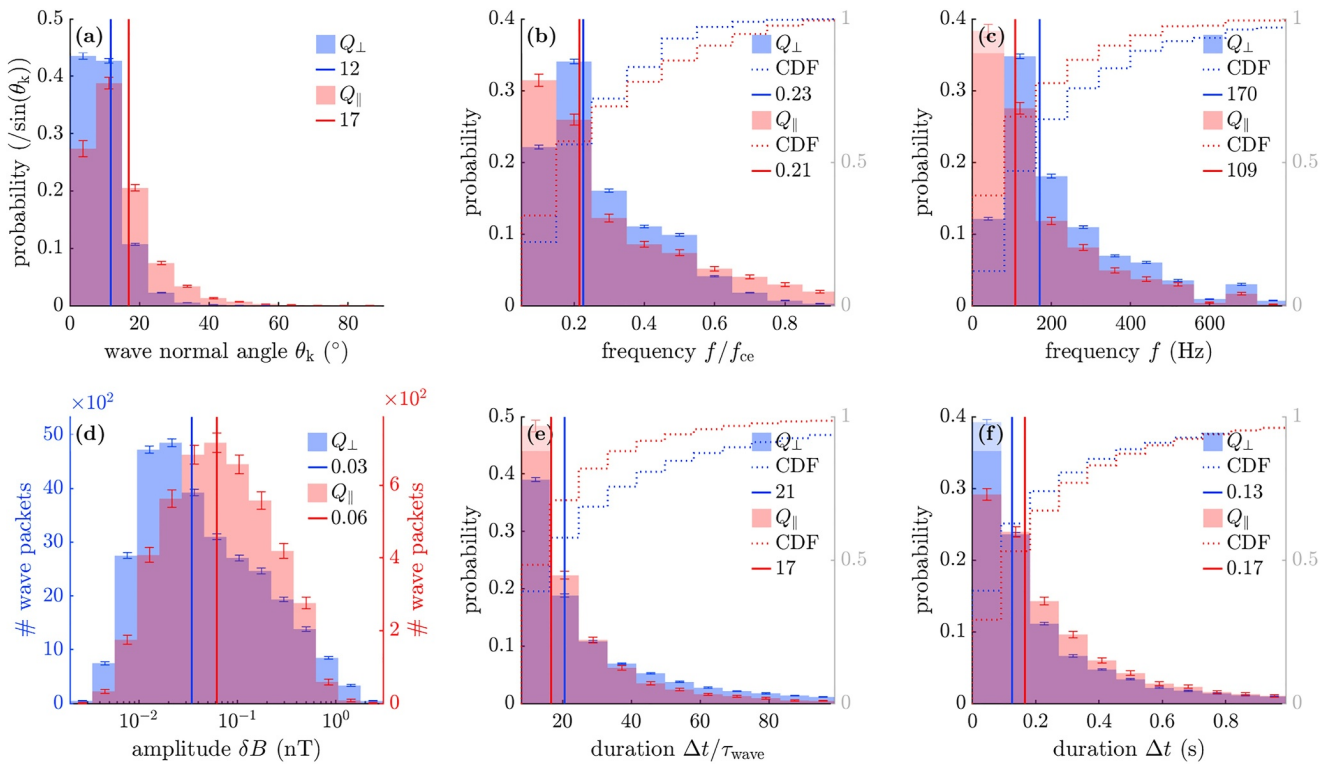

**Figure 4.** Properties of whistler waves in the $Q_\perp$ (blue) and $Q_\parallel$ (red) MSH. (a) Wave normal angle $\theta_k$ (probability normalized to $\sin(\theta_k)$ which would correspond to a uniform distribution of wave vector directions). (b) Normalized wave frequency $f/f_{ce}$. (c) Wave frequency $f$ in Hz. (d) Amplitude $\delta B$. (e) Duration $\Delta t$ normalized to the wave period $\tau_{wave}$. (f) $\Delta t$ in seconds. The vertical red and blue lines indicate median values in the $Q_\perp$ and $Q_\parallel$ MSH, respectively, with values given in the legend. The cumulative distributions (CDFs; dotted) are measured on the right y-scale (gray). The error bars are estimated by the standard Poisson error $\sqrt{N}$, where $N$ is the counts in each bin.

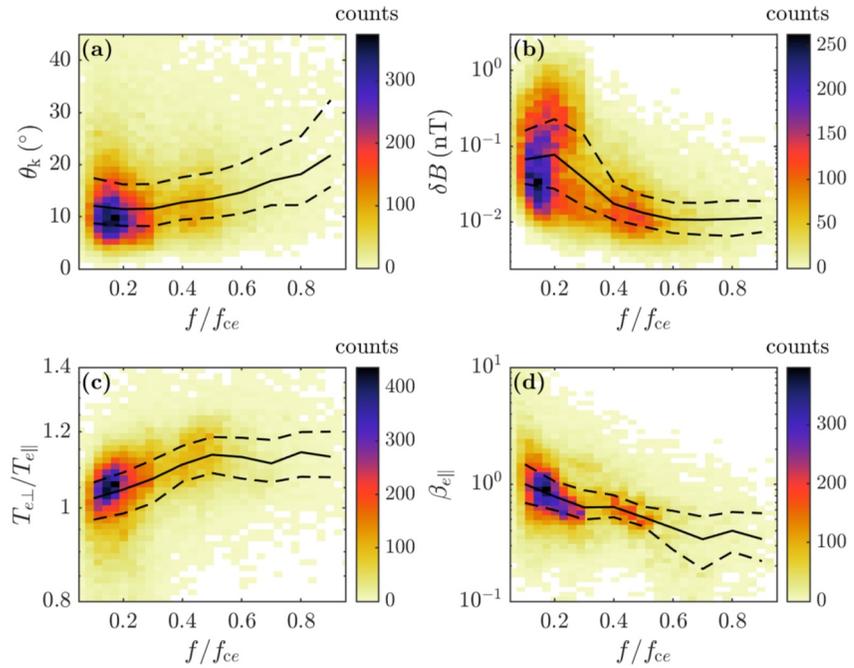

**Figure 5.** 2D histograms of key wave properties. We show $f/f_{ce}$ versus (a) $\theta_k$; (b) $\delta B$; (c) and (d) time average of $T_{e\perp}/T_{e\parallel}$ and $\beta_{e\parallel}$ during each wave packet. The solid lines show the median in bins centered at $f/f_{ce} = 0.1, 0.2, \ldots, 0.9$. The dashed lines show the 25th–75th percentiles.

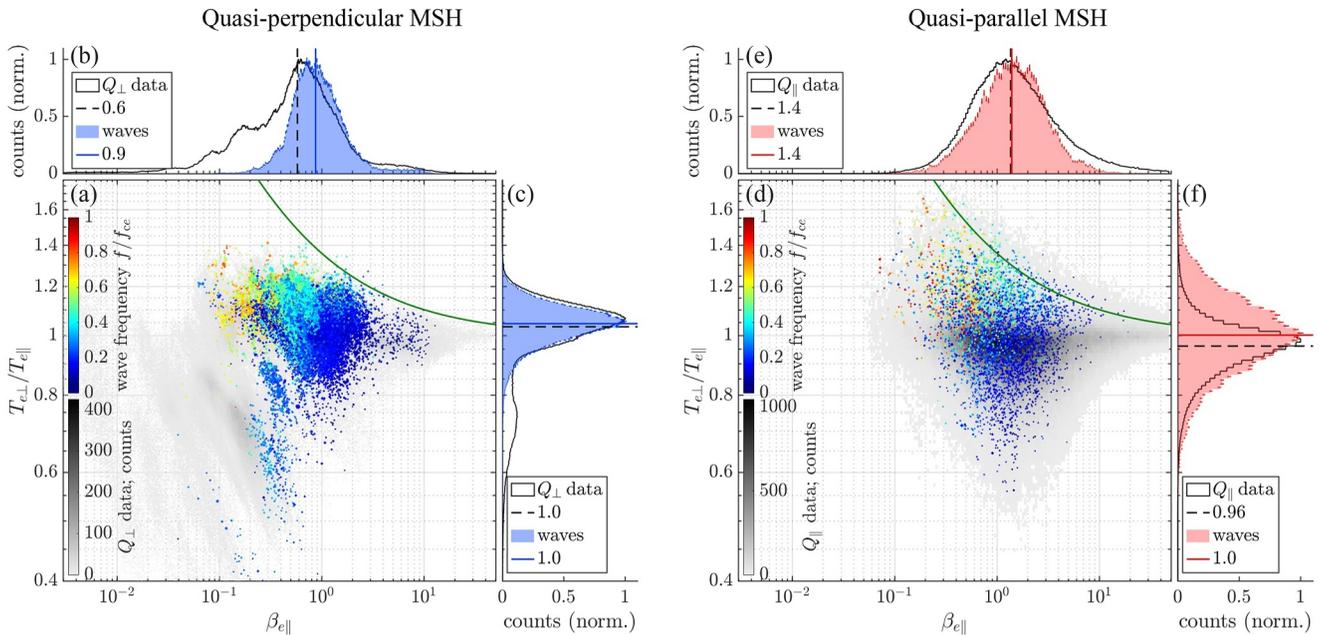

**Figure 6.** Distribution of data and wave packets in the $\beta_{e\parallel}$-$T_{e\perp}/T_{e\parallel}$ parameter space. (a) 2D histogram of the $Q_\perp$ data intervals in greyscale. The overplotted points are whistler wave packets with color showing their normalized frequency $f/f_{ce}$ (plotting the time average of $T_{e\perp}/T_{e\parallel}$ and $\beta_{e\parallel}$ during each wave packet). (b) Distribution of $\beta_{e\parallel}$ in the $Q_\perp$ MSH (black curve, median shown with a black dashed line) and during wave packets (blue filled, median shown with a blue line). (c) Distribution of $T_{e\perp}/T_{e\parallel}$ in the $Q_\perp$ MSH (black curve, median shown with a black dashed line) and during wave packets (blue filled, median shown with a blue line). (d)–(f) Same as (a)–(c) but for the $Q_\parallel$ MSH. In (b), (c), (e), and (f), the counts have been normalized to their maximum values. The error bars are estimated by the standard Poisson error $\sqrt{N}$, where $N$ is the counts in each bin. The green curves in (a) and (d) show the whistler instability threshold (Gary & Wang, 1996) for $\gamma = 0.01|\Omega_{ce}|$.

Figure 6a shows the distribution of wave packets and plasma measurements in the $\beta_{e\parallel}$-$T_{e\perp}/T_{e\parallel}$ space. The gray background shows the 2D histogram of the whole $Q_\perp$ MSH data set. The left part, $\beta_{e\parallel} < 0.1$, is caused by the disturbed solar wind intervals mentioned above. In the upper right, the data is approximately limited by the whistler instability threshold (green curve), indicating that whistler waves play a role in regulating the plasma in the $Q_\perp$ MSH. The individual colored points represent the average $\beta_{e\parallel}$ and $T_{e\perp}/T_{e\parallel}$ values during each whistler wave packet. The color code shows $f/f_{ce}$. As for the whole $Q_\perp$ MSH data set, the vast majority of the wave packets are detected in regions predicted to be stable to the whistler instability. The waves are almost absent from the regions with $\beta_{e\parallel} < 0.1$. We mainly observe waves with $f < 0.3f_{ce}$ (dark blue), but higher frequencies up to $f_{ce}$ are also found. The higher-frequency waves tend to occur in lower-$\beta_{e\parallel}$ and higher-$T_{e\perp}/T_{e\parallel}$ regions compared to lower-frequency waves, consistent with Figures 5c and 5d.

In the $Q_\parallel$ MSH, we find the median $\beta_{e\parallel} = 1.4$, which is higher than in the $Q_\perp$ MSH, while the median $T_{e\perp}/T_{e\parallel} = 1.0$ is similar for the two data sets (Figures 6e and 6f). The $Q_\parallel$ MSH data set lacks the low-$\beta_{e\parallel}$ tail (compare panels b and e) since it does not include any disturbed solar wind intervals. There is no significant difference in $\beta_{e\parallel}$ and $T_{e\perp}/T_{e\parallel}$ between the whole $Q_\parallel$ MSH data set and the times when whistler waves are observed. The 2D histogram for the $Q_\parallel$ MSH is shown in Figure 6d. Several features of the distribution are similar to what we found in the $Q_\perp$ MSH: most of the data is limited by the whistler instability threshold, most waves are observed in regions predicted to be stable to the whistler instability, and the higher-frequency waves tend to appear in regions with low $\beta_{e\parallel}$ and high $T_{e\perp}/T_{e\parallel}$.

In summary, in both the $Q_\perp$ and $Q_\parallel$ MSH, there is no significant difference in the median $T_{e\perp}/T_{e\parallel}$ when whistler waves are observed compared to the background plasma (i.e., the whole $Q_\perp$ and $Q_\parallel$ data sets). In the $Q_\perp$ MSH, we observe an increased median $\beta_{e\parallel}$ during wave observations. In both MSH geometries, most waves observations were made where the plasma is predicted to be stable to the whistler instability, assuming a bi-Maxwellian electron distribution. In the $Q_\perp$ MSH, our data set extends to very low $\beta_{e\parallel}$ values of $\sim 10^{-2}$, and we found that whistlers are unlikely to appear in regions with $\beta_{e\parallel} < 0.1$.

### 3.3. Correlation With Magnetic Field and Density Structures

Whistler waves have previously been reported to occur within mirror mode waves in the $Q_\perp$ MSH (Kitamura et al., 2020; Smith & Tsurutani, 1976), and more generally in local magnetic minima in the $Q_\parallel$ MSH (Svenningsson et al., 2022). We now use our large data set of whistlers in both MSH geometries to investigate if this is the general case. In Figures 7a–7d, we show examples of whistler waves from our data set, observed in the $Q_\perp$ MSH. Throughout the interval, we observe whistlers as magnetic power enhancements (Figure 7a) between $0.05f_{ce}$ and $f_{ce}$ (white curves). Some of these whistler emissions, for example, at 15:19:18, 15:20:15, 15:21:25, and 15:21:55, are associated with bi-directional Poynting flux (see panel b). This indicates wave propagation in directions both parallel and anti-parallel to $\boldsymbol{B}$ and is typically interpreted as a signature of the spacecraft crossing a wave source region (Y. Zhang et al., 1998). In Figures 7c and 7d, the solid curves show $B$ and $n_i$, respectively, and the dashed curves, $B_{\mathrm{filt}}$ and $n_{i,\mathrm{filt}}$, show the same signals low-pass filtered below the frequency $f_{\mathrm{filt}} = 1/t_{\mathrm{filt}}$ where $t_{\mathrm{filt}} = 60$ s. We observe that the whistler waves in this interval tend to occur in the regions of $B < B_{\mathrm{filt}}$ (magnetic dips) and $n_i > n_{i,\mathrm{filt}}$ (density peaks).

We now investigate the whistler occurrence in magnetic and density dips and peaks for the entire whistler database. Since the sizes of magnetic and density structures range between different scales (mirror modes, for example, have been observed on time scales 3.5–24 s (Soucek et al., 2008)), we compute the occurrence using eight different values of $t_{\mathrm{filt}}$, ranging from 1 to 120 s. Figure 7e shows the distribution of $B - B_{\mathrm{filt}}$ in the $Q_\perp$ MSH, boxes indicating the 25th–75th percentiles and horizontal lines showing the medians. For each time scale $t_{\mathrm{filt}}$, gray represents all the data, darker blue corresponds to the whistler wave packets, and lighter blue to wave packets with amplitudes exceeding 0.1 nT (around 30% of all wave packets, see Figure 4d). The numbers above the box plots show the corresponding fraction of all wave data found inside the magnetic dips. The numbers in parenthesis show the same but for the $\delta B > 0.1$ nT waves. We see that the whistler waves are preferentially observed in magnetic dips and that the preference is the strongest (80%) for high-amplitude waves using the longest $t_{filt}$ of 120 s. There is also a preference toward density peaks (66% for $\delta B > 0.1$ nT, $t_{filt} = 120$, see Figure 7f). Corresponding plots for the $Q_\parallel$ MSH are shown in Figures 7g and 7h. Again, the whistler waves are preferentially observed in magnetic dips and density peaks, and the relation is more pronounced for high-amplitude (>0.1 nT) waves. However, the preference is weaker in the $Q_\parallel$ MSH (70% for magnetic dips and 60% for density peaks) compared to the $Q_\perp$ MSH. The time scales with the strongest preference are 5–30 s, shorter than in the $Q_\perp$ MSH.

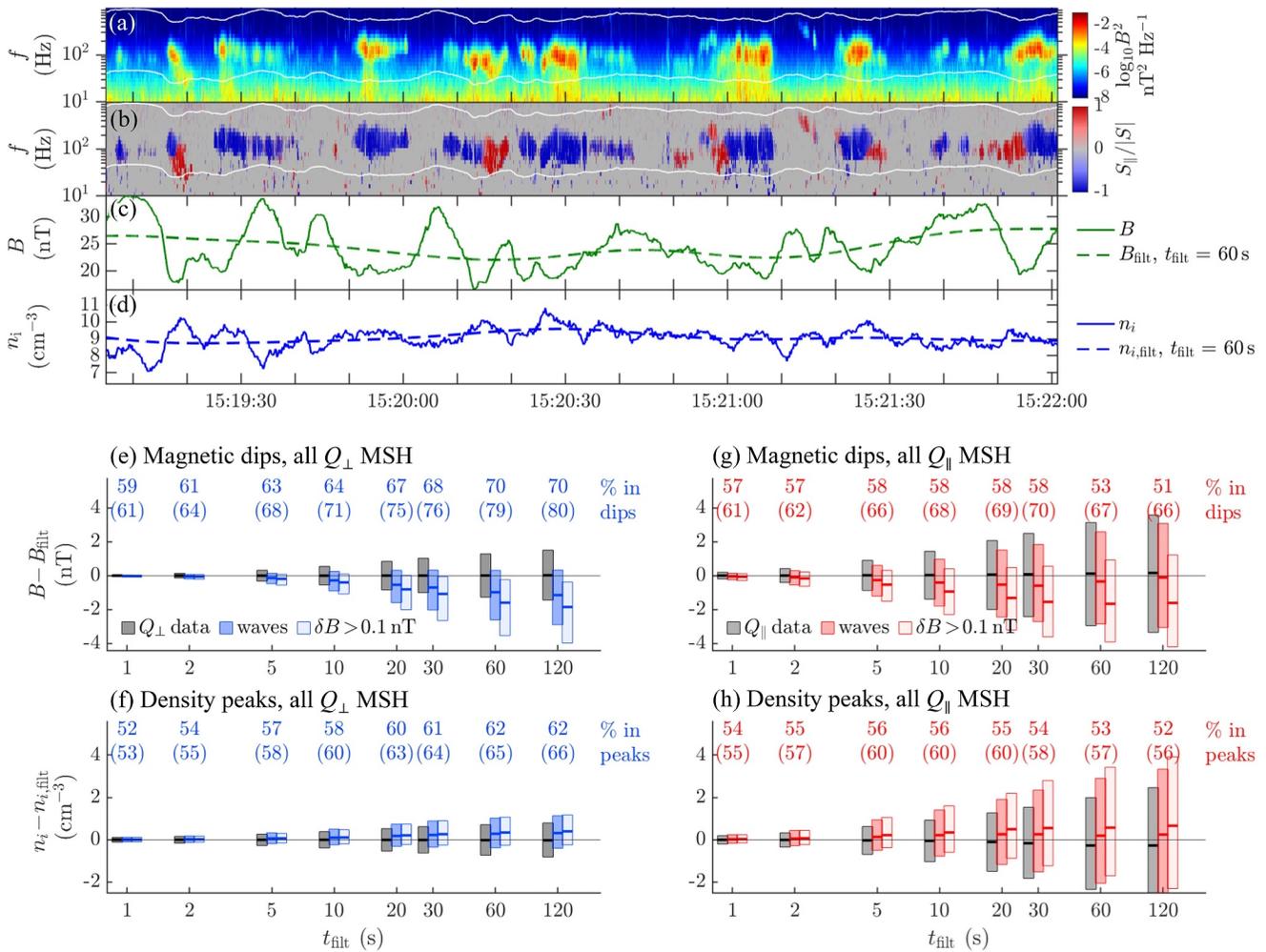

**Figure 7.** Whistler occurrence in magnetic minima and density peaks. (a)–(d): Example of a 3-min interval in the $Q_\perp$ MSH (2023-03-06) with whistler waves observed in magnetic dips and density peaks. (a) Magnetic power spectrum. The white curves show $0.05f_{ce}$ and $f_{ce}$, between which we detect whistler waves. Whistlers are observed throughout the interval. (b) Poynting flux direction. (c) Magnetic field magnitude $B$, unfiltered (green solid) and low-pass filtered (green dashed). (d) Ion density $n_i$, unfiltered (blue solid) and low-pass filtered (blue dashed). For this example $t_{filt} = 60$ s. (e)–(h): Distributions for the entire whistler database. (e): Distribution of $B - B_{filt}$ in the $Q_\perp$ MSH, boxes showing the 25th–75th percentiles and horizontal lines showing the medians. For each time scale $t_{filt}$, gray represents all data and blue only intervals during whistler observations. Light-blue boxes show waves with $\delta B > 0.1$ nT. The numbers above each $t_{filt}$ show the percentage of time spent in magnetic dips ($B < B_{filt}$) for all waves (without parenthesis), and the $\delta B > 0.1$ nT waves (in parenthesis). (f): Distribution of $n_i - n_{i,filt}$ in the $Q_\perp$ MSH, same format as (e). (g), (h) Same as (e), (f) but in the $Q_\parallel$ MSH, red color representing wave intervals.

In summary, across the different MSH geometries and time scales, we find that the whistlers are associated with a lower $B$ and higher $n_i$ than the surroundings. In addition, we note that in both MSH geometries, the preference toward $B$ dips is stronger than that toward $n_i$ peaks.

## 4. Discussion

In this section, we first discuss our observations in Section 4.1. Then, in Section 4.2, we estimate the effect of pitch-angle scattering from the waves. Finally, in Section 4.3, we discuss how the resonant energy relates to $\beta_{e\parallel}$ and $f/f_{ce}$ in our wave observations.

### 4.1. Occurrence and Wave Properties

In both MSH geometries, we found that whistler waves tend to propagate quasi-parallel to the background magnetic field. We also found that whistler waves are preferentially observed in local magnetic dips, consistent with previous whistler observations in the $Q_\perp$ MSH (Smith & Tsurutani, 1976), and density peaks. Since this preference was stronger for the high-amplitude waves, the whistler waves are likely generated inside these structures (as waves escaping the source region are likely to be damped by the surrounding stable plasma). Local wave generation is also motivated by the bi-directional Poynting flux observed in the example in Figure 7b. Parallel propagation and occurrence in local magnetic dips are consistent with case studies of whistler waves in the MSH (Kitamura et al., 2020; Svenningsson et al., 2022; H. Zhang et al., 2021). In these works, similar generation mechanisms were proposed: the local dip in $B$ creates an electron velocity distribution (trapped at pitch angles $\theta$ between $\theta_{tr}$ and $180° - \theta_{tr}$) with local velocity-space minima parallel and anti-parallel to the background magnetic field. This distribution contains the free energy to cause growth of whistler waves, mainly through first-order cyclotron resonance ($l = 1$, Equation 3) which produces parallel propagating waves with frequency $0.05$–$0.25f_{ce}$. This suggests that the observed whistler waves with frequencies around $0.05$–$0.25f_{ce}$ are generated in local magnetic dips.

We can estimate the dominant scales of the $B$ dips and $n$ peaks that are relevant for the whistler waves by considering the occurrence of whistler waves inside dips/peaks as a function of the filter frequency $f_{filt} = 1/t_{filt}$. The upper bound for the size of the resolved peaks is roughly $t_{filt}/2$, that is, half a wavelength corresponding to $f_{filt}$. The occurrence of whistlers inside dips/peaks is the highest for $t_{filt}$ above 30 s in the $Q_\perp$ MSH, indicating that structures of the order of some tens of seconds are the most impactful for whistler generation. In the $Q_\parallel$ MSH, where the correlation between whistlers and these structures is weaker, the same analysis yields scales of a few seconds. With a typical bulk speed of 200 km/s this corresponds to scales of $\geq 1,000$ km in the $Q_\perp$ MSH and $100$–$2,000$ km in the $Q_\parallel$ MSH. Since the ion gyroradius in the MSH is about 100 km (using $T_i = 500$ eV, $B = 30$ nT), we find that the whistler-associated structures in the $Q_\parallel$ MSH structures reach down to ion scales, while the corresponding structures in the $Q_\perp$ MSH are slightly larger.

The higher whistler occurrence in the $Q_\perp$ MSH (see Table 1) shows that this geometry provides more favorable conditions for whistler wave generation. This is not surprising since mirror modes, known to act as whistler source regions (Kitamura et al., 2020), are more common in the $Q_\perp$ MSH (Dimmock et al., 2015). Among whistler source regions that have been described in previous works, most have been of ion scale sizes or larger. Yordanova et al. (2020) reported that strong magnetic fluctuations on electron-to-ion scales are common in the $Q_\parallel$ MSH, while almost absent in the $Q_\perp$ MSH. The more stationary nature and the absence of small-scale fluctuations in the $Q_\perp$ MSH likely provide good conditions for whistler source regions to form and persist, leading to a higher occurrence in this geometry. Nonetheless, we have shown that whistler waves are indeed present in the $Q_\parallel$ as well (filling factor of 6%, see Table 1) and that they can be more intense than those in the $Q_\perp$ MSH (see Figure 4d). Overall, however, the wave properties ($\theta_k$, $f/f_{ce}$, $\Delta t$) are similar in the two geometries, which indicates that similar whistler generation mechanisms are at play in the $Q_\perp$ and $Q_\parallel$ MSH.

In Figure 6, we found that the whistler waves in the MSH are not associated with a high local temperature anisotropy: the mean $T_{e\perp}/T_{e\parallel}$ was ~1 in both the $Q_\perp$ and $Q_\parallel$ MSH, independent of the presence of whistler waves. This could in principle, to some extent, be due to the waves propagating away from their source regions. However, it is also known that electron distributions in the MSH are non-Maxwellian (Graham et al., 2021) and that local features in the distribution function can be responsible for wave growth even if $T_{e\perp}/T_{e\parallel}$ is close to 1. Several non-Maxwellian features can affect $T_{e\perp}/T_{e\parallel}$ and the stability of the electron distribution. The thermal part of the distribution is often dominated by a flat top (Feldman et al., 1983) which has more parallel-flowing electrons at low energies. A better correlation between whistler waves and $T_{e\perp}/T_{e\parallel}$ may therefore be obtained by restricting the energy range to the resonant energy of the waves (see Figure 8c) (Breuillard, Matteini, et al., 2018). However, there also exist distributions with low anisotropy at resonant energies which have been found unstable to whistler generation in the MSH, such as the butterfly distribution (Kitamura et al., 2020; Svenningsson et al., 2022). Non-

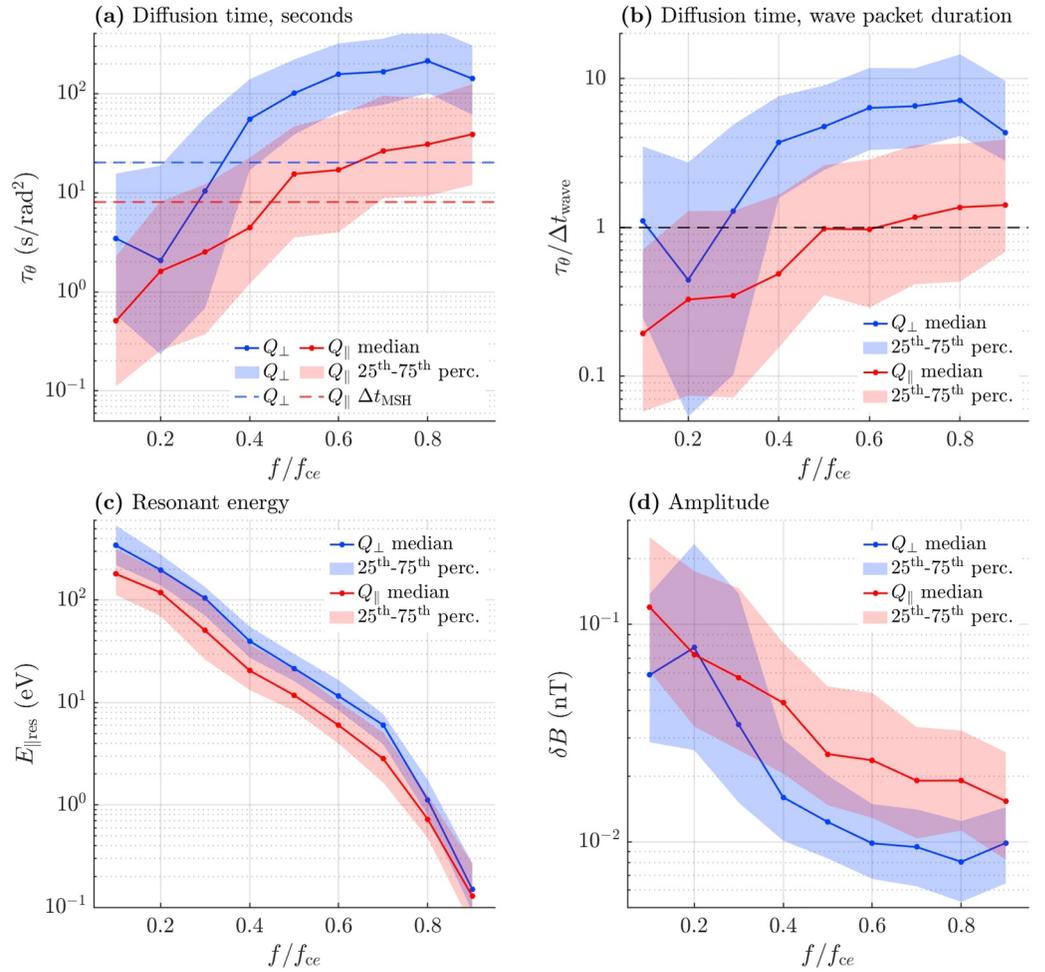

**Figure 8.** (a) Pitch-angle diffusion time $\tau_\theta$ in the $Q_\perp$ (blue) and $Q_\parallel$ (red) MSH. The dots show the median in bins centered at $f/f_{ce} = 0.1, 0.2, ..., 0.9$. The shaded areas indicate the 25th and 75th percentiles. The dashed blue and red lines show the whistler interaction time $\Delta t_{MSH}$ in the $Q_\perp$ and $Q_\parallel$ MSH, respectively. (b) Same as (a), but with $\tau_\theta$ normalized to wave packet duration $\Delta t_{wave}$. (c) Resonant energy $E_{\parallel res}$ in eV. (d) Wave amplitude $\delta B$ in nT.

Maxwellian velocity distributions such as those mentioned above may partly explain the high whistler occurrence in regions predicted stable by $T_{e\perp}/T_{e\parallel}$.

## 4.2. Estimation of the Pitch-Angle Diffusion Time

To estimate the effect of the observed waves on the electron velocity distribution, we calculate the pitch-angle diffusion coefficients for the observed wave packets. While an exact quantification of whistler-associated diffusion is not achievable statistically, we can get an estimation of their importance for the electron dynamics. Whistler waves interact with the part of the electron distribution satisfying the cyclotron resonance condition:

$$k_\parallel v_{\parallel res} = \omega - l|\Omega_{ce}|, \qquad l = 0, \pm 1, \pm 2, ... \qquad (3)$$

where $k_\parallel$ is the whistler wave vector component parallel to $\boldsymbol{B}$, $v_\parallel$ the parallel electron velocity, and $\omega$ the angular frequency. For simplicity, we restrict the analysis to parallel propagating waves, which is approximately the case for the observed waves (see Figure 4a). For parallel waves, only the first-order resonance ($l = +1$) contributes to velocity-space diffusion (Verscharen et al., 2022). We use the cold plasma dispersion relation for parallel-propagating whistler waves,

$$\frac{k_{\parallel}^2 c^2}{\omega^2} = 1 - \frac{\omega_{pi}^2}{\omega(\omega + |\Omega_{ci}|)} - \frac{\omega_{pe}^2}{\omega(\omega - |\Omega_{ce}|)}, \tag{4}$$

to calculate $k_{\parallel}$ and the group velocity $d\omega/dk_{\parallel}$ (Kennel & Petschek, 1966), where $\omega_{pi,e} = \sqrt{n_{i,e} q_{i,e}^2/(m_{i,e} \epsilon_0)}$ and $\Omega_{ci,e} = q_{i,e} B/m_{i,e}$ are the angular plasma and cyclotron frequencies computed using $B$ and $n_e$ averaged over the wave packet duration, and $c$ is the speed of light. The interaction of whistlers with electrons results in velocity-space diffusion along surfaces that are locally tangent to circles centered at the phase velocity $v_{ph} = \omega/k_{\parallel}$ (Kennel & Engelmann, 1966):

$$v_{\perp}^2 + (v_{\parallel} - v_{ph})^2 = \text{constant.} \tag{5}$$

The diffusion smooths out velocity-space gradients along this surface around the resonant velocity $v_{\parallel \text{res}}$. This reduces the anisotropy around the resonant energy in the case of wave growth (see Shi et al. (2020) for observations in the Earth's foreshock), and scatters electrons toward larger $v_{\perp}$ in the case of wave damping (as shown in a simulation by Jeong et al. (2020)).

For this analysis, we focus on the low-frequency waves, $\omega \lesssim 0.3|\Omega_{ce}|$, which corresponds to the majority of the observed wave packets. It follows from Equations 5 and 3 that when $\omega \ll |\Omega_{ce}|$, $|v_{ph,i}| \ll |v_{\parallel \text{res}}|$, and the diffusion surfaces resemble surfaces of constant energy. Thus, interaction with low-frequency waves mainly changes the pitch-angle of electrons while their energy is almost unchanged. Under the additional assumption that the electrons are randomly distributed in phase with respect to the waves, the evolution of the electron velocity distribution $f_e$ can be described by the reduced Fokker-Planck equation (Kennel & Petschek, 1966):

$$\frac{\partial f_e}{\partial t} = \frac{1}{\sin\theta} \frac{\partial}{\partial\theta}\left(D_\theta \sin\theta \frac{\partial f_e}{\partial\theta}\right), \tag{6}$$

where $D_\theta$ is the pitch-angle diffusion coefficient:

$$D_\theta = \left(\frac{q_e}{m_e}\right)^2 \frac{\delta B^2}{\Delta k} \frac{1}{v_{\parallel}} \approx \left(\frac{q_e}{m_e}\right)^2 \frac{\delta B^2}{2\pi\Delta f} \frac{1}{v_{\parallel \text{res}}} \frac{d\omega}{dk}. \tag{7}$$

Equation 7 holds for narrow-band waves such that $\delta B^2/\Delta k \approx (d\omega/dk)\delta B^2/\Delta\omega$.

We calculate $D_\theta$ for all 34,409 observed wave packets, using $\delta B^2/\Delta f = 2P_{\max}^w$ where $P_{\max}^w$ is the maximum value of the power spectral density (see Section 2.2) and $v_{\parallel \text{res}}$ from Equation 3 to estimate the diffusion time scale with $\tau_\theta = 1/D_\theta$. In general, the change in $f_e$ also depends on the shape of $f_e$ through the pitch-angle derivative in Equation 6. In Appendix A, we evaluate Equation 6 and show that $f_e (\partial f_e/\partial t)^{-1}$ is less than an order of magnitude away from $1/D_\theta$ for our wave packets, and therefore $\tau_\theta = 1/D_\theta$ provides a reasonable approximation for the diffusion time scale.

The blue and red curves in Figure 8a show the median $\tau_\theta$ in the $Q_\perp$ and $Q_\parallel$ MSH, respectively, and shaded regions give the 25th to 75th percentile. We observe $\tau_\theta$ values ranging from 0.1 to 1,000 s/rad$^2$, and differences are mainly caused by variations in wave amplitudes (see panel d). The lower-frequency waves, $f \sim 0.1$–$0.2 f_{ce}$, tend to have a higher amplitude and therefore cause stronger diffusion; in this range, approximately 75% of the waves have $\tau_\theta < 10$ s/rad$^2$. The higher-frequency waves, $f \gtrsim 0.4 f_{ce}$, have a significantly shorter diffusion time in the $Q_\parallel$ MSH as they have higher amplitude compared to the $Q_\perp$ geometry (see Figure 8d).

We now want to establish whether the pitch-angle diffusion is strong enough to significantly reshape the electron velocity distribution, in other words, if the time spent interacting with the waves while crossing the MSH is enough to allow diffusion to influence the electron population. In the $Q_\perp$ MSH, the median $\tau_\theta$ is longer than the wave packet duration $\Delta t_{\text{wave}}$ at most frequencies (see Figure 8b; blue), indicating that most whistler packets are too short to significantly reshape $f_e$. However, in 40% of the wave packets, $\tau_\theta < \Delta t_{\text{wave}}$, in which case the wave causes significant pitch-angle diffusion in $f_e$. Around 20% have $\tau_\theta < 0.1\Delta t_{\text{wave}}$, indicating a strong effect on $f_e$. In the $Q_\parallel$ MSH, $\tau_\theta/\Delta t_{wave}$ is smaller (see Figure 8b; red). Here, 70% of the wave packets have $\tau_\theta < \Delta t_{wave}$ and 30% have $\tau_\theta < 0.1\Delta t_{wave}$. We conclude that 40% of the whistlers in the $Q_\perp$ and 70% of the whistlers in the $Q_\parallel$ MSH are strong enough to cause significant pitch-angle scattering during the wave packet duration. We recall that the change in $f_e$ occurs around the resonant energy $E_{\parallel res} = m_e v_{\parallel res}^2/2$, which is around 100 eV for $f \approx 0.2 f_{ce}$ (see Figure 8c).

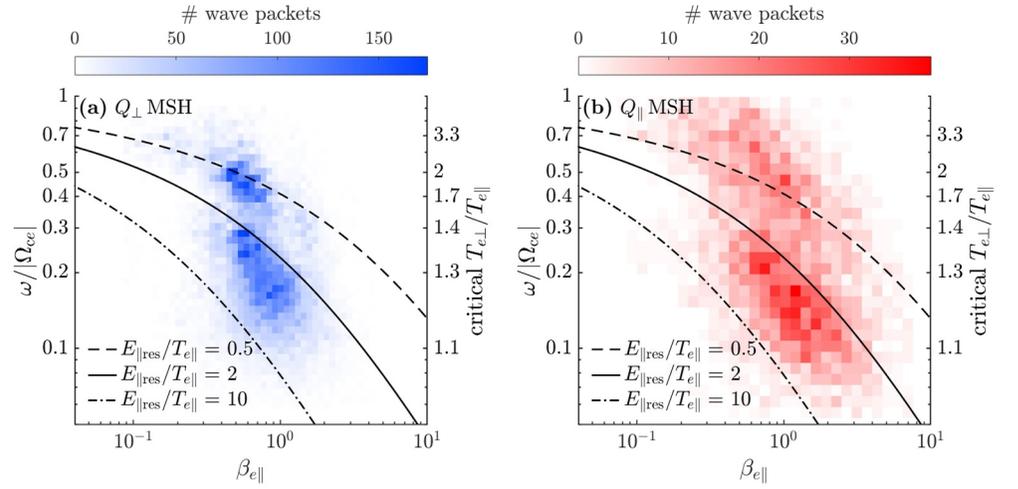

**Figure 9.** Whistler occurrence in the $\beta_{e\parallel}$-$\omega/|\Omega_{ce}|$ parameter space, 2D histograms of wave packets using the time average of $\beta_{e\parallel}$ during each wave packet. Black curves show Equation 8 for three values of $E_{\parallel res}/T_{e\parallel}$ when first-order cyclotron resonance is assumed. The right $y$ axes show the critical anisotropy needed for whistler generation (Kennel & Petschek, 1966).

A plasma parcel traveling across the magnetosheath can encounter multiple wave packets. We compare $\tau_\theta$ to the interaction time $\Delta t_{MSH}$, which is the time a plasma parcel spends interacting with whistlers in the MSH. With a typical speed of 200 km/s and MSH thickness of 3–5 $R_E$, we estimate a transit time of 95–160 s. Further, the whistler filling factor (see Table 1) gives the fraction of that crossing during which the plasma will encounter whistler waves. We estimate the effective interaction time by multiplying the transit time with the filling factor (17% in the $Q_\perp$ and 6% in the $Q_\parallel$ MSH), obtaining $\Delta t_{MSH} \approx 10$–20 s for the $Q_\perp$ MSH and $\Delta t_{MSH} \approx 5$–8 s in the $Q_\parallel$ MSH (dashed lines in Figure 8a). We find significant pitch-angle diffusion, that is, $\tau_\theta < \Delta t_{MSH}$, for 60% of the waves in the $Q_\perp$ MSH and for 70% of the waves in the $Q_\parallel$ MSH. We conclude that the combined effect of whistler wave encounters can alter the shape of the electron velocity distribution in both MSH geometries.

We note that for the higher-frequency whistlers, the pitch-angle diffusion is weaker, mainly due to their lower amplitude. However, since the low-frequency limit cannot be applied to the analysis of the higher-frequency wave packets, the energy diffusion can be more important in this case, for example, causing strong damping of the waves.

### 4.3. Resonant Energy and $\beta_{e\parallel}$ Lower Limit

In this section, we use the resonant electron energy to discuss the distribution of our identified wave packets in $\beta_{e\parallel}$ and wave frequency. Using the resonance condition (Equation 3) together with the cold plasma dispersion relation for parallel whistler waves (Equation 4), we calculate the parallel energy $E_{\parallel res} = m_e v_{\parallel res}^2/2$ of electrons in first-order cyclotron resonance with the waves (Kennel & Petschek, 1966):

$$\frac{E_{\parallel res}}{T_{e\parallel}} = \frac{1}{\beta_{e\parallel}} \frac{|\Omega_{ce}|}{\omega} \left(1 - \frac{\omega}{|\Omega_{ce}|}\right)^3. \tag{8}$$

The black curves in Figure 9 show $\omega/|\Omega_{ce}|$ as a function of $\beta_{e\parallel}$ calculated from Equation 8 for different values of $E_{\parallel res}/T_{e\parallel}$. The color scales show 2D histograms for the observed whistler wave packets. In both the $Q_\perp$ (blue; Figure 9a) and $Q_\parallel$ MSH (red; Figure 9b), we observe two populations of waves; one with lower frequency $(0.05 \lesssim \omega/|\Omega_{ce}| \lesssim 0.3)$ and one with higher frequency ($\omega/|\Omega_{ce}| \gtrsim 0.4$). Both populations are slightly tilted so that higher frequencies occur at lower $\beta_{e\parallel}$.

As shown in Figure 6, the whistler observations are almost absent in regions with $\beta_{e\parallel} \lesssim 0.1$. This can be explained by considering the limits on resonant energy $E_{res\parallel}$ and frequency imposed by the first-order cyclotron resonance:

1. *Resonant energy.* Since the phase-space density drops with increasing energy, $E_{\parallel res}/T_{e\parallel}$ can not be too high in order to have sufficiently many electrons available to interact with the waves. For example, at $E_{\parallel res} = 10T_{e\parallel}$ (dash-dotted curves in Figure 9a and 9b), a basic computation for a Maxwellian distribution yields a phase space density $10^{-4}$ times that at $E_{\parallel res} = T_{e\parallel}$. Below/left of this curve, whistler resonance is ineffective due to a lack of resonant electrons.
2. *Frequency.* An upper limit in frequency comes from the critical anisotropy needed for whistler instability (assuming a bi-Maxwellian electron distribution). The anisotropy required for wave generation increases with frequency through $T_{e\perp}/T_{e\parallel} > 1 + 1/(|\Omega_{ce}|/\omega - 1)$ (Kennel & Petschek, 1966). For example, if $\omega = 0.3|\Omega_{ce}|$, a bi-Maxwellian would need a temperature anisotropy of $T_{e\perp}/T_{e\parallel} > 1.4$ to lead to whistler growth instead of damping. This is higher than most $T_{e\perp}/T_{e\parallel}$ values in the MSH (see Figure 6).

With these considerations, if we require for example, $\omega < 0.3|\Omega_{ce}|$ and $E_{\parallel res} < 10T_{e\parallel}$, we get the lower limit on $\beta_{e\parallel}$ of ~0.1, consistent with the observed wave distributions in Figures 6 and 9. The lower-frequency waves $(0.05 \lesssim \omega/|\Omega_{ce}| \lesssim 0.3)$ have $E_{res\parallel} \sim 2$–$5T_{e\parallel}$ and sufficiently low frequency to be consistent with generation through first-order cyclotron resonance.

The higher-frequency whistler waves likely have a different generation mechanism compared to the low-frequency ones. For example, with higher-order resonances (i.e. Equation 3 with $l = -1$ or $l = 2$), the $\omega \sim 0.5|\Omega_{ce}|$ waves could access resonant energies on the order of 2–5$T_{e\parallel}$. In addition, since the resonant velocity would be higher, the critical anisotropy would decrease. A future study will be needed to determine the source of the higher-frequency whistler waves.

## 5. Conclusions

We have statistically studied the occurrence and properties of whistler waves in the Earth's magnetosheath (MSH), comparing the quasi-perpendicular ($Q_\perp$) and quasi-parallel ($Q_\parallel$) MSH geometries. We identified 34,409 whistler waves using an automatic search and characterized their properties including frequency, duration, amplitude, and wave normal angle.

The whistler wave occurrence is higher in the $Q_\perp$ MSH with a filling factor 17% in comparison to 6% in the $Q_\parallel$ MSH. In both geometries, the whistler waves have a median normalized frequency $f/f_{ce} = 0.2$. The median duration is 0.13 s in the $Q_\perp$ and 0.17 s in the $Q_\parallel$ MSH, corresponding to a few tens of wave periods in both cases. They propagate quasi-parallel to the background magnetic field (median $\theta_k = 12°$ in the $Q_\perp$ and 17° in the $Q_\parallel$ MSH). Their median amplitude in the $Q_\parallel$ MSH is 0.06 nT (or 60 pT), higher than in the $Q_\perp$ MSH (0.03 nT or 30 pT).

We find that in both MSH geometries, whistler waves are preferentially observed in magnetic dips and density peaks. The strongest preference is for waves with amplitudes above 0.1 nT in the $Q_\perp$ MSH, where 80% of the wave activity is found in magnetic dips and 66% in density peaks. Correspondingly, the maximum in the $Q_\parallel$ MSH was 70% in magnetic dips and 60% in density peaks. The whistler waves are associated with smaller-scale structures (1–10 s) in the $Q_\parallel$ MSH. In the $Q_\perp$ MSH, they are associated with larger (minute-scale) structures.

We very rarely observe whistler waves in plasma regions with $\beta_{e\parallel} < 0.1$, and we speculate that this is due to the resonant energy becoming too high compared to the electron thermal energy so that there are not enough resonant electrons to drive the waves. The presence of whistler waves is not associated with a high electron temperature anisotropy, contrary to what is expected from the temperature anisotropy instability. We suggest that this is in part explained by non-Maxwellian features in the electron distribution.

An important result is that for low/intermediate-frequency whistlers ($f/f_{ce} < 0.3$), the pitch-angle diffusion time is less than the time a typical plasma parcel interacts with whistlers in the MSH, indicating that wave-particle interaction with whistler waves can significantly modify the electron velocity distribution in both the $Q_\perp$ and $Q_\parallel$ MSH.

## Appendix A: Pitch-Angle Diffusion

In Section 4.2, we approximated the diffusion time scale with $\tau_\theta = 1/D_\theta$. Below, we will evaluate the diffusion equation to discuss the accuracy of this approximation.

### A1. Examples of Diffusion Paths

Figure A1 shows examples of diffusion in different electron distributions interacting with whistler waves of different normalized frequencies ($\omega/\Omega_{ce}| = (0.13, 0.23, 0.45)$, the same values that we use in Section A2). We assume that the wave vector is anti-parallel to **B**, resulting in $v_{ph} < 0$ and $v_{res\parallel} > 0$ (see Equation 3, $l = 1$). Electrons in resonance with whistler waves diffuse along velocity-space curves locally tangent to circles centered at $v_{ph}$ (black arrows; see Equation 5) around $v_{res\parallel}$ (vertical black dashed line). Since the diffusion works to smooth out gradients in $f_e$, it is directed from high to low values of $f_e$. In Figures A1a–A1c, $f_e$ is Maxwellian (i.e., isotropic, $T_{e\perp}/T_{e\parallel} = 1$), so the isocontours of $f_e$ (white) coincide with curves of constant energy (blue). As a consequence, for a Maxwellian, the particle diffusion is always toward higher energy, increasing the total energy in the distribution. With the lowest frequency, $\omega/\Omega_{ce}| = 0.13$ (Figure A1a), $v_{\parallel res}$ is large compared to $v_{ph}$, and the diffusion curves are almost aligned with circles of constant energy. This means that pitch-angle diffusion dominates. There is, however, a weak gradient along the diffusion curve which leads to a weak particle diffusion (black arrows) toward higher energy, that is, wave damping. When going to higher frequencies ($\omega/\Omega_{ce}| = 0.23$ and 0.45; Figures A1b and A1c), $v_{\parallel res}$ (vertical black dashed line) approaches $v_{ph}$ (vertical black dash-dotted line) and the diffusion curves deviate more from curves of constant energy. This results in a stronger gradient along the diffusion curves and more wave damping. Figure A1d shows a bi-Maxwellian with $T_{e\perp}/T_{e\parallel} = 1.5$. The high anisotropy reverses the direction of the diffusion (black arrows) toward lower particle energy, leading to wave growth.

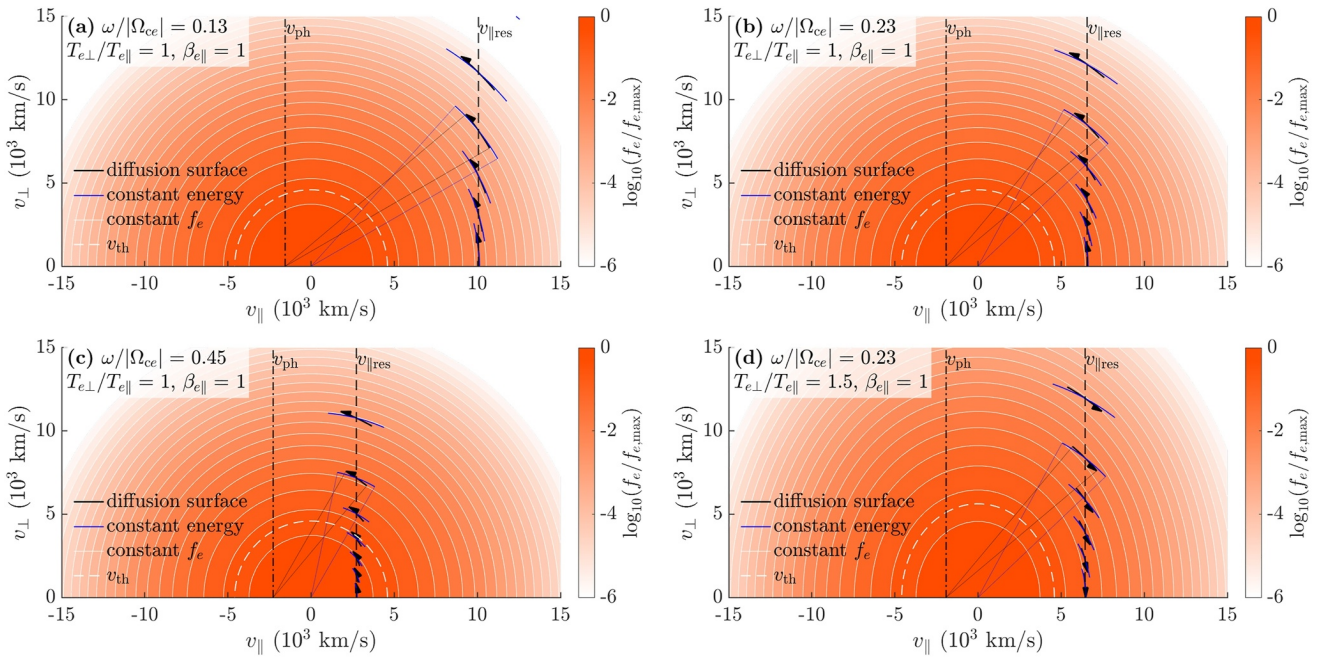

**Figure A1.** Diffusion curves for first-order cyclotron resonance. The black dash-dotted and dashed lines show $v_\parallel = v_{ph}$ and $v_\parallel = v_{res\parallel}$, respectively. The white solid curves show isocontours of $f_e$ and the white dashed curve shows the thermal velocity. Diffusion curves are shown in black (Equation 5), with arrows pointing in the direction of the diffusion (toward lower $f_e$ values). Blue indicates curves of constant energy (for readability only plotted around $v_{\parallel res}$).

### A2. Diffusion Time Approximation

To investigate the rate of change in $f_e$ during the whistler interaction, we solve Equation 6, which is formally valid in the $\omega \ll |\Omega_{ce}|$ limit, where $|v_{res\parallel}| \gg |v_{ph}|$. In this limit, at velocities around $v_{res\parallel}$, the pitch angle in the wave frame ($\theta_{wave}$) is the same as the pitch angle in the electron bulk frame ($\theta_e$). When evaluating the derivatives in

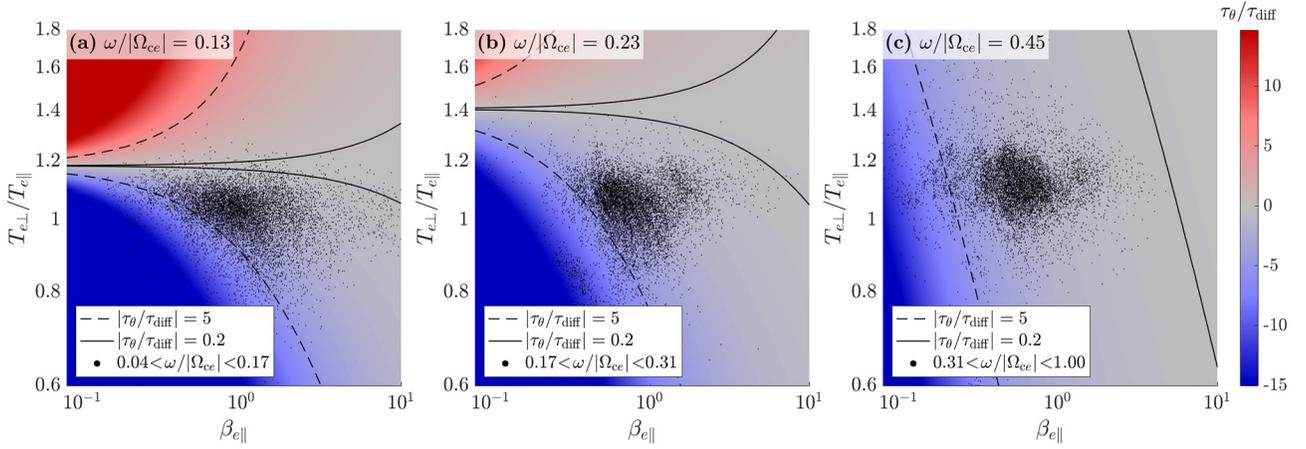

**Figure A2.** Diffusion time correction factor $\tau_\theta/\tau_{\mathrm{diff}}$ (Equation A2) in the $\beta_{e\parallel}$-$T_{e\perp}/T_{e\parallel}$ parameter space for different normalized wave frequencies: $\omega/|\Omega_{ce}| = 0.13$ (a), 0.23 (b), and 0.45 (c). The dashed and solid curves show isocontours for $|\tau_\theta/\tau_{\mathrm{diff}}| = 5$ and 0.2, respectively. The black points in each subplot show all wave packets in a frequency interval containing the chosen $\omega$.

Equation 6, we use $\theta = \theta_{\mathrm{wave}}$ which gives the gradients along the diffusion surfaces. We define the diffusion time scale as

$$\tau_{\mathrm{diff}} = \frac{f_e}{\partial f_e/\partial t}\bigg|_{v=v_{\mathrm{res}\parallel},\,\theta\approx0}.$$  (A1)

Since $D_\theta$ (Equation 7) is independent of $\theta$, Equation 6 becomes

$$\frac{\partial f_e}{\partial t} = D_\theta \frac{1}{\sin\theta} \frac{\partial}{\partial\theta}\left(\sin\theta \frac{\partial f_e}{\partial\theta}\right) \quad \Rightarrow \quad \frac{\tau_\theta}{\tau_{\mathrm{diff}}} = \frac{1}{f_e}\left[\frac{1}{\sin\theta}\frac{\partial}{\partial\theta}\left(\sin\theta\frac{\partial f_e}{\partial\theta}\right)\right],$$  (A2)

where $\tau_\theta = 1/D_\theta$ is the diffusion time used in Section 4.2. We use Equation A2 to evaluate how well $\tau_\theta$ can approximate $\tau_{\mathrm{diff}}$. We consider a bi-Maxwellian $f_e$ with the fixed density $n_e = 15\ \mathrm{cm}^{-3}$ and parallel temperature $T_{e\parallel} = 60\ \mathrm{eV}$, which are typical values found in the MSH. We create gradients in $f_e$ ($\partial f_e/\partial\theta$ in Equation A2) by varying the temperature anisotropy $T_{e\perp}/T_{e\parallel}$. It is important to remember that electron velocity distributions are rarely Maxwellian or bi-Maxwellian in the MSH and that the actual gradients in the distributions could differ from what is assumed from the temperature anisotropy. The value of $T_{e\perp}/T_{e\parallel}$ in this analysis should be interpreted as a proxy for velocity-space gradients rather than the moment-calculated anisotropy of the full distribution.

In Figure A2, we plot $\tau_\theta/\tau_{\mathrm{diff}}$ from Equation A2 in the $\beta_{e\parallel}$-$T_{e\perp}/T_{e\parallel}$ parameter space for three frequencies: $\omega/|\Omega_{ce}| = 0.13$ (panel a), 0.23 (panel b), and 0.45 (panel c). To cover values of $\beta_{e\parallel}$ and $T_{e\perp}/T_{e\parallel}$ corresponding to our whistler observations, we sweep $B$ between 6 and 150 nT and $T_{e\perp}$ from 23 to 108 eV. Negative $\tau_{\mathrm{diff}}$ (blue), caused by $\partial f_e/\partial\theta < 0$, correspond to increased particle energy or wave damping (such as in Figures A1a–A1c) while positive $\tau_{\mathrm{diff}}$ (red) correspond to reduced particle energy or wave growth (such as in Figure A1d). The black dots represent the whistler wave packets in frequency ranges chosen so that the number of wave packets is the same in each interval. The center frequencies are the median in each interval. The dashed and solid curves show isocontours of $|\tau_\theta/\tau_{\mathrm{diff}}| = 5$ and 0.2, respectively. When $|\tau_\theta/\tau_{\mathrm{diff}}|$ is large (left of the dashed curves), the diffusion time scale $\tau_{\mathrm{diff}}$ is shorter, that is, the diffusion is stronger, than what is implied by $\tau_\theta$. Between the dashed and solid curves, $|\tau_{\mathrm{diff}}|$ is less than a factor of 5 away from $\tau_\theta$. The majority of our wave packets (black dots) are found inside this interval. This motivates the use of $\tau_\theta$ as an approximation for $\tau_{\mathrm{diff}}$ in Section 4.2.

## Data Availability Statement

Magnetospheric multiscale data are available at https://lasp.colorado.edu/mms/sdc/public/data/ following the directories: mms#/fgm/brst/l2 for FGM data, mms#/scm/brst/l2 for SCM data, mms#/edp/brst/l2 for EDP data,

mms#/fpi/brst/l2/dis-dist for FPI ion distributions, mms#/fpi/brst/l2/dis-moms for FPI ion moments mms#/fpi/brst/l2/des-dist for FPI electron distributions, and mms#/fpi/brst/l2/des-moms for FPI electron moments. Data analysis was performed using the IRFU-Matlab analysis package. No new data has been produced as part of this project.

**Acknowledgments**
We thank the entire MMS team and instrument PIs for data access and support. This work is supported by the Swedish Research Council Grant 2016-0550 and the Swedish National Space Agency Grant 158/16. EY is supported by the Swedish National Space Agency Grant 145/18 and 192/20.